
\font \eightbf         = cmbx8
\font \eighti          = cmmi8
\font \eightit         = cmti8
\font \eightrm         = cmr8
\font \eightsl         = cmsl8
\font \eightsy         = cmsy8
\font \eighttt         = cmtt8
\font \tenbf           = cmbx9
\font \teni            = cmmi9
\font \tenit           = cmti9
\font \tenrm           = cmr9
\font \tensl           = cmsl9
\font \tensy           = cmsy9
\font \tentt           = cmtt9

\font \kleinhalbcurs   = cmmib10 scaled 800

\font \sixbf           = cmbx6
\font \sixi            = cmmi6
\font \sixrm           = cmr6
\font \sixsy           = cmsy6
\font \tafonts         = cmbx12
\font \tafontss        = cmbx10
\font \tafontt         = cmbx10 scaled\magstep2
\font \tams            = cmmib10
\font \tenmib          = cmmib10
\font \tamt            = cmmib10
\font \tass            = cmsy10
\font \tasss           = cmsy7
\font \tast            = cmsy10 scaled\magstep2
\font \tbfonts         = cmbx8
\font \tbfontss        = cmbx10  scaled 667
\font \tbfontt         = cmbx10 scaled\magstep1
\font \tbmt            = cmmib10
\font \tbss            = cmsy8
\font \tbsss           = cmsy6
\font \tbst            = cmsy10  scaled\magstep1
\vsize=23.5truecm
\hoffset=-1true cm
\voffset=-1true cm
\newdimen\fullhsize
\fullhsize=40cc
\hsize=19.5cc
\def\fullline{\hbox to\fullhsize}
\def\makefootline{\baselineskip=10dd \fullline{\the\footline}}
\def\makeheadline{\vbox to 0pt{\vskip-22.5pt
            \fullline{\vbox to 8.5pt{}\the\headline}\vss}\nointerlineskip}
\let\lr=L \newbox\leftcolumn
\output={\global\topskip=10pt
         \if L\lr
            \global\setbox\leftcolumn=\columnbox \global\let\lr=R
            \message{[left\the\pageno]}%
            \ifnum\pageno=1
               \global\topskip=\fullhead\fi
         \else
            \doubleformat \global\let\lr=L
         \fi
         \ifnum\outputpenalty>-2000 \else\dosupereject\fi}
\def\doubleformat{\shipout\vbox{\makeheadline
    \fullline{\box\leftcolumn\hfil\columnbox}
           \makefootline}
           \advancepageno}
\def\columnbox{\leftline{\pagebody}}
\outer\def\bye{\bigskip\typeset
\sterne=1\ifx\speciali\undefined\else
\loop\smallskip\noindent special character No\number\sterne:
\csname special\romannumeral\sterne\endcsname
\advance\sterne by 1\global\sterne=\sterne
\ifnum\sterne<11\repeat\fi
\if R\lr\null\fi\vfill\supereject\end}
\def\typeset{\begpet\noindent This article was processed by the author using
Sprin\-ger-Ver\-lag \TeX\ AA macro package 1989.\endpet}
\hfuzz=2pt
\vfuzz=2pt
\tolerance=1000
\fontdimen3\tenrm=1.5\fontdimen3\tenrm
\fontdimen7\tenrm=1.5\fontdimen7\tenrm
\abovedisplayskip=3 mm plus6pt minus 4pt
\belowdisplayskip=3 mm plus6pt minus 4pt
\abovedisplayshortskip=0mm plus6pt
\belowdisplayshortskip=2 mm plus4pt minus 4pt
\predisplaypenalty=0
\clubpenalty=20000
\widowpenalty=20000
\parindent=1.5em
\frenchspacing
\def\newline{\hfill\break}%
\nopagenumbers
\def\AALogo{\setbox254=\hbox{ ASTROPHYSICS }%
\vbox{\baselineskip=10dd\hrule\hbox{\vrule\vbox{\kern3pt
\hbox to\wd254{\hfil ASTRONOMY\hfil}
\hbox to\wd254{\hfil AND\hfil}\copy254
\hbox to\wd254{\hfil\number\day.\number\month.\number\year\hfil}
\kern3pt}\vrule}\hrule}}
\def\paglay{\headline={{\tenrm\hsize=.75\fullhsize\ifnum\pageno=1
\vbox{\baselineskip=10dd\hrule\line{\vrule\kern3pt\vbox{\kern3pt
\hbox{\bf A and A Manuskript-Nr.}
\hbox{(will be inserted by hand later)}
\kern3pt\hrule\kern3pt
\hbox{\bf Your thesaurus codes are:}
\hbox{\rightskip=0pt plus3em\advance\hsize by-7pt
\vbox{\noindent\ignorespaces\the\THESAURUS}}
\kern3pt}\hfil\kern3pt\vrule}\hrule}
\rlap{\quad\AALogo}\hfil
\else\ifodd\pageno\hfil\folio\else\folio\hfil\fi\fi}}}
\ifx \undefined\instruct
\headline={\tenrm\ifodd\pageno\hfil\folio
\else\folio\hfil\fi}\fi
\newcount\eqnum\eqnum=0%
\def\autnum{\global\advance\eqnum by 1{\rm(\the\eqnum)}}
\newtoks\eq\newtoks\eqn
\catcode`@=11
\def\eqalign#1{\null\vcenter{\openup\jot\m@th
  \ialign{\strut\hfil$\displaystyle{##}$&$\displaystyle{{}##}$\hfil
      \crcr#1\crcr}}}
\def\displaylines#1{{}$\displ@y
\hbox{\vbox{\halign{$\@lign\hfil\displaystyle##\hfil$\crcr
    #1\crcr}}}${}}
\def\eqalignno#1{{}$\displ@y
  \hbox{\vbox{\halign to\hsize{\hfil$\@lign\displaystyle{##}$\tabskip\z@skip
    &$\@lign\displaystyle{{}##}$\hfil\tabskip\centering
    &\llap{$\@lign##$}\tabskip\z@skip\crcr
    #1\crcr}}}${}}
\def\leqalignno#1{{}$\displ@y
\hbox{\vbox{\halign
to\hsize{\qquad\hfil$\@lign\displaystyle{##}$\tabskip
\z@skip
    &$\@lign\displaystyle{{}##}$\hfil\tabskip\centering
    &\kern-\hsize\rlap{$\@lign##$}\tabskip\hsize\crcr
    #1\crcr}}}${}}
\def\generaldisplay{%
\ifeqno
       \ifleqno\leftline{$\displaystyle\the\eqn\quad\the\eq$}%
       \else\line{$\displaystyle\the\eq\hfill\the\eqn$}\fi
\else
       \leftline{$\displaystyle\the\eq$}%
\fi
\global\eq={}\global\eqn={}}%
\newif\ifeqno\newif\ifleqno \everydisplay{\displaysetup}
\def\displaysetup#1$${\displaytest#1\eqno\eqno\displaytest}
\def\displaytest#1\eqno#2\eqno#3\displaytest{%
\if!#3!\ldisplaytest#1\leqno\leqno\ldisplaytest
\else\eqnotrue\leqnofalse\eqn={#2}\eq={#1}\fi
\generaldisplay$$}
\def\ldisplaytest#1\leqno#2\leqno#3\ldisplaytest{\eq={#1}%
\if!#3!\eqnofalse\else\eqnotrue\leqnotrue\eqn={#2}\fi}
\catcode`@=12 %
\mathchardef\Gamma="0100
\mathchardef\Delta="0101
\mathchardef\Theta="0102
\mathchardef\Lambda="0103
\mathchardef\Xi="0104
\mathchardef\Pi="0105
\mathchardef\Sigma="0106
\mathchardef\Upsilon="0107
\mathchardef\Phi="0108
\mathchardef\Psi="0109
\mathchardef\Omega="010A

\def\utw{\smash{\rlap{\lower5pt\hbox{$\sim$}}}}
\def\udtw{\smash{\rlap{\lower6pt\hbox{$\approx$}}}}

\def\diameter{{\ifmmode\mathchoice
{\ooalign{\hfil\hbox{$\displaystyle/$}\hfil\crcr
{\hbox{$\displaystyle\mathchar"20D$}}}}
{\ooalign{\hfil\hbox{$\textstyle/$}\hfil\crcr
{\hbox{$\textstyle\mathchar"20D$}}}}
{\ooalign{\hfil\hbox{$\scriptstyle/$}\hfil\crcr
{\hbox{$\scriptstyle\mathchar"20D$}}}}
{\ooalign{\hfil\hbox{$\scriptscriptstyle/$}\hfil\crcr
{\hbox{$\scriptscriptstyle\mathchar"20D$}}}}
\else{\ooalign{\hfil/\hfil\crcr\mathhexbox20D}}%
\fi}}

\normallineskip=1dd
\normallineskiplimit=0dd
\normalbaselineskip=10dd
\textfont0=\tenrm
\textfont1=\teni
\textfont2=\tensy
\textfont\itfam=\tenit
\textfont\slfam=\tensl
\textfont\ttfam=\tentt
\textfont\bffam=\tenbf
\normalbaselines\rm
\def\petit{\def\rm{\fam0\eightrm}%
\textfont0=\eightrm \scriptfont0=\sixrm \scriptscriptfont0=\fiverm
 \textfont1=\eighti \scriptfont1=\sixi \scriptscriptfont1=\fivei
 \textfont2=\eightsy \scriptfont2=\sixsy \scriptscriptfont2=\fivesy
 \def\it{\fam\itfam\eightit}%
 \textfont\itfam=\eightit
 \def\sl{\fam\slfam\eightsl}%
 \textfont\slfam=\eightsl
 \def\bf{\fam\bffam\eightbf}%
 \textfont\bffam=\eightbf \scriptfont\bffam=\sixbf
 \scriptscriptfont\bffam=\fivebf
 \def\tt{\fam\ttfam\eighttt}%
 \textfont\ttfam=\eighttt
 \let\tams=\kleinhalbcurs
 \let\tenbf=\eightbf
 \let\sevenbf=\sixbf
 \normalbaselineskip=9dd
 \if Y\REFEREE \normalbaselineskip=2\normalbaselineskip
 \normallineskip=2\normallineskip\fi
 \setbox\strutbox=\hbox{\vrule height7pt depth2pt width0pt}%
 \normalbaselines\rm}%
\def\begpet{\vskip6pt\bgroup\petit}%
\def\endpet{\vskip6pt\egroup}%
\def\rahmen#1{\vbox{\hrule\line{\vrule\vbox to#1true
cm{\vfil}\hfil\vrule}\vfil\hrule}}
\def\begfig#1cm#2\endfig{\par
   \ifvoid\topins\midinsert\bigskip\vbox{\rahmen{#1}#2}\endinsert
   \else\topinsert\vbox{\rahmen{#1}#2}\endinsert%
\fi}
\def\begfigwid#1cm#2\endfig{\par
\if N\lr\else
\if R\lr
\shipout\vbox{\makeheadline
\line{\box\leftcolumn}\makefootline}\advancepageno
\fi\let\lr=N
\topskip=10pt
\output={\plainoutput}%
\fi
\topinsert\line{\vbox{\hsize=\fullhsize\rahmen{#1}#2}\hss}\endinsert}
\def\figure#1#2{\bigskip\noindent{\petit{\bf Fig.\ts#1.\
}\ignorespaces #2\smallskip}}
\def\begtab#1cm#2\endtab{\par
   \ifvoid\topins\midinsert\medskip\vbox{#2\rahmen{#1}}\endinsert
   \else\topinsert\vbox{#2\rahmen{#1}}\endinsert%
\fi}
\def\begtabemptywid#1cm#2\endtab{\par
\if N\lr\else
\if R\lr
\shipout\vbox{\makeheadline
\line{\box\leftcolumn}\makefootline}\advancepageno
\fi\let\lr=N
\topskip=10pt
\output={\plainoutput}%
\fi
\topinsert\line{\vbox{\hsize=\fullhsize#2\rahmen{#1}}\hss}\endinsert}
\def\begtabfullwid#1\endtab{\par
\if N\lr\else
\if R\lr
\shipout\vbox{\makeheadline
\line{\box\leftcolumn}\makefootline}\advancepageno
\fi\let\lr=N
\output={\plainoutput}%
\fi
\topinsert\line{\vbox{\hsize=\fullhsize\noindent#1}\hss}\endinsert}

\def\begref{\vskip1cm\begingroup\let\INS=N}
\def\ref{\goodbreak\if N\INS\let\INS=Y\vbox{\noindent\tenbf
References\bigskip}\fi\hangindent\parindent
\hangafter=1\noindent\ignorespaces}
\def\endref{\goodbreak\endgroup}%
\def\ack#1{\vskip11pt\begingroup\noindent{\it Acknowledgements\/}.
\ignorespaces#1\vskip6pt\endgroup}

 \def \aTa  { \goodbreak
     \bgroup
     \par
 \textfont0=\tafontt \scriptfont0=\tafonts \scriptscriptfont0=\tafontss
 \textfont1=\tamt \scriptfont1=\tbmt \scriptscriptfont1=\tams
 \textfont2=\tast \scriptfont2=\tass \scriptscriptfont2=\tasss
     \baselineskip=17dd
     \lineskip=17dd
     \rightskip=0pt plus2cm\spaceskip=.3333em \xspaceskip=.5em
     \pretolerance=10000
     \noindent
     \tafontt}
 \def \eTa{\vskip10pt\egroup
     \noindent
     \ignorespaces}

 \def \aTb{\goodbreak
     \bgroup
     \par
 \textfont0=\tbfontt \scriptfont0=\tbfonts \scriptscriptfont0=\tbfontss
 \textfont1=\tbmt \scriptfont1=\tenmib \scriptscriptfont1=\tams
 \textfont2=\tbst \scriptfont2=\tbss \scriptscriptfont2=\tbsss
     \baselineskip=13dd
     \lineskip=13dd
     \rightskip=0pt plus2cm\spaceskip=.3333em \xspaceskip=.5em
     \pretolerance=10000
     \noindent
     \tbfontt}
 \def \eTb{\vskip10pt
     \egroup
     \noindent
     \ignorespaces}
\catcode`\@=11
\expandafter \newcount \csname c@Tl\endcsname
    \csname c@Tl\endcsname=0
\expandafter \newcount \csname c@Tm\endcsname
    \csname c@Tm\endcsname=0
\expandafter \newcount \csname c@Tn\endcsname
    \csname c@Tn\endcsname=0
\expandafter \newcount \csname c@To\endcsname
    \csname c@To\endcsname=0
\expandafter \newcount \csname c@Tp\endcsname
    \csname c@Tp\endcsname=0
\def \resetcount#1    {\global
    \csname c@#1\endcsname=0}
\def\@nameuse#1{\csname #1\endcsname}
\def\arabic#1{\@arabic{\@nameuse{c@#1}}}
\def\@arabic#1{\ifnum #1>0 \number #1\fi}
\expandafter \newcount \csname c@fn\endcsname
    \csname c@fn\endcsname=0
\def \stepc#1    {\global
    \expandafter
    \advance
    \csname c@#1\endcsname by 1}
\catcode`\@=12
   \catcode`\@= 11

\skewchar\eighti='177 \skewchar\sixi='177
\skewchar\eightsy='60 \skewchar\sixsy='60
\hyphenchar\eighttt=-1
\def\footnoterule{\kern-3pt\hrule width 2true cm\kern2.6pt}%
\newinsert\footins
\def\footnotea#1{\let\@sf\empty %
  \ifhmode\edef\@sf{\spacefactor\the\spacefactor}\/\fi
  {#1}\@sf\vfootnote{#1}}
\def\vfootnote#1{\insert\footins\bgroup
  \textfont0=\tenrm\scriptfont0=\sevenrm\scriptscriptfont0=\fiverm
  \textfont1=\teni\scriptfont1=\seveni\scriptscriptfont1=\fivei
  \textfont2=\tensy\scriptfont2=\sevensy\scriptscriptfont2=\fivesy
  \interlinepenalty\interfootnotelinepenalty
  \splittopskip\ht\strutbox %
  \splitmaxdepth\dp\strutbox \floatingpenalty\@MM
  \leftskip\z@skip \rightskip\z@skip \spaceskip\z@skip \xspaceskip\z@skip
  \textindent{#1}\footstrut\futurelet\next\fo@t}
\def\fo@t{\ifcat\bgroup\noexpand\next \let\next\f@@t
  \else\let\next\f@t\fi \next}
\def\f@@t{\bgroup\aftergroup\@foot\let\next}
\def\f@t#1{#1\@foot}
\def\@foot{\strut\egroup}
\def\footstrut{\vbox to\splittopskip{}}
\skip\footins=\bigskipamount %
\count\footins=1000 %
\dimen\footins=8in %
   \def \bfootax  {\bgroup\tenrm
                  \baselineskip=12pt\lineskiplimit=-6pt
                  \hsize=19.5cc
                  \def\textindent##1{\hang\noindent\hbox
                  to\parindent{##1\hss}\ignorespaces}%
                  \footnotea{$^\star$}\bgroup}
   \def \efootax  {\egroup\egroup}
   \def \bfootay  {\bgroup\tenrm
                  \baselineskip=12pt\lineskiplimit=-6pt
                  \hsize=19.5cc
                  \def\textindent##1{\hang\noindent\hbox
                  to\parindent{##1\hss}\ignorespaces}%
                  \footnotea{$^{\star\star}$}\bgroup}
   \def \efootay  {\egroup\egroup }
   \def \bfootaz {\bgroup\tenrm
                  \baselineskip=12pt\lineskiplimit=-6pt
                  \hsize=19.5cc
                  \def\textindent##1{\hang\noindent\hbox
                  to\parindent{##1\hss}\ignorespaces}%
                 \footnotea{$^{\star\star\star}$}\bgroup}
   \def \efootaz {\egroup \egroup}
\def\fonote#1{\mehrsterne$^{\the\sterne}$\begingroup
       \def\textindent##1{\hang\noindent\hbox
       to\parindent{##1\hss}\ignorespaces}%
\vfootnote{$^{\the\sterne}$}{#1}\endgroup}
\catcode`\@=12
\everypar={\let\lasttitle=N\everypar={\parindent=1.5em}}%
\def \titlea#1{\stepc{Tl}
     \resetcount{Tm}
     \vskip22pt
     \setbox0=\vbox{\vskip 22pt\noindent
     \bf
     \rightskip 0pt plus4em
     \pretolerance=20000
     \arabic{Tl}.\
   \textfont1=\tams\scriptfont1=\kleinhalbcurs\scriptscriptfont1=
\kleinhalbcurs
     \ignorespaces#1
     \vskip11pt}
     \dimen0=\ht0\advance\dimen0 by\dp0\advance\dimen0 by 2\baselineskip
     \advance\dimen0 by\pagetotal
     \ifdim\dimen0>\pagegoal\eject\fi
     \bgroup
     \noindent
     \bf
     \rightskip 0pt plus4em
     \pretolerance=20000
     \arabic{Tl}.\
\textfont1=\tams\scriptfont1=\kleinhalbcurs\scriptscriptfont1=
\kleinhalbcurs
     \ignorespaces#1
     \vskip11pt
     \egroup
     \nobreak
     \parindent=0pt
     \everypar={\parindent=1.5em
     \let\lasttitle=N\everypar={\let\lasttitle=N}}%
     \let\lasttitle=A%
     \ignorespaces}
 \def\titleb#1{\stepc{Tm}
     \resetcount{Tn}
     \if N\lasttitle\else\vskip-11pt\vskip-\baselineskip
     \fi
     \vskip 17pt
     \setbox0=\vbox{\vskip 17pt
     \raggedright
     \pretolerance=10000
     \noindent
     \it
     \arabic{Tl}.\arabic{Tm}.\
     \ignorespaces#1
     \vskip8pt}
     \dimen0=\ht0\advance\dimen0 by\dp0\advance\dimen0 by 2\baselineskip
     \advance\dimen0 by\pagetotal
     \ifdim\dimen0>\pagegoal\eject\fi
     \bgroup
     \raggedright
     \pretolerance=10000
     \noindent
     \it
     \arabic{Tl}.\arabic{Tm}.\
     \ignorespaces#1
     \vskip8pt
     \egroup
     \nobreak
     \let\lasttitle=B%
     \parindent=0pt
     \everypar={\parindent=1.5em
     \let\lasttitle=N\everypar={\let\lasttitle=N}}%
     \ignorespaces}
 \def \titlec#1{\stepc{Tn}
     \resetcount{To}
     \if N\lasttitle\else\vskip-3pt\vskip-\baselineskip
     \fi
     \vskip 11pt
     \setbox0=\vbox{\vskip 11pt
     \noindent
     \raggedright
     \pretolerance=10000
     \arabic{Tl}.\arabic{Tm}.\arabic{Tn}.\
     \ignorespaces#1\vskip6pt}
     \dimen0=\ht0\advance\dimen0 by\dp0\advance\dimen0 by 2\baselineskip
     \advance\dimen0 by\pagetotal
     \ifdim\dimen0>\pagegoal\eject\fi
     \bgroup\noindent
     \raggedright
     \pretolerance=10000
     \arabic{Tl}.\arabic{Tm}.\arabic{Tn}.\
     \ignorespaces#1\vskip6pt
     \egroup
     \nobreak
     \let\lasttitle=C%
     \parindent=0pt
     \everypar={\parindent=1.5em
     \let\lasttitle=N\everypar={\let\lasttitle=N}}%
     \ignorespaces}
 \def\titled#1{\stepc{To}
     \resetcount{Tp}
     \if N\lasttitle\else\vskip-3pt\vskip-\baselineskip
     \fi
     \vskip 11pt
     \bgroup
     \it
     \noindent
     \ignorespaces#1\unskip. \egroup\ignorespaces}
\let\REFEREE=N
\newbox\refereebox
\setbox\refereebox=\vbox
to0pt{\vskip0.5cm\fullline{\hrulefill\tentt\lower0.5ex
\hbox{\kern5pt referee's copy\kern5pt}\hrulefill}\vss}%
\def\refereelayout{\let\REFEREE=M\footline={\copy\refereebox}%
\message{|A referee's copy will be produced}\par
\if N\lr\else
\if R\lr
\shipout\vbox{\makeheadline
\line{\box\leftcolumn}\makefootline}\advancepageno
\fi\let\lr=N
\topskip=10pt
\output={\plainoutput}%
\fi
}
\let\ts=\thinspace
\newcount\sterne \sterne=0
\newdimen\fullhead
\newtoks\RECDATE
\newtoks\ACCDATE
\newtoks\MAINTITLE
\newtoks\SUBTITLE
\newtoks\AUTHOR
\newtoks\INSTITUTE
\newtoks\SUMMARY
\newtoks\KEYWORDS
\newtoks\THESAURUS
\newtoks\SENDOFF
\newlinechar=`\| %
\catcode`\@=\active
\let\INS=N%
\def@#1{\if N\INS $^{#1}$\else\if
E\INS\hangindent0.5\parindent\hangafter=1%
\noindent\hbox to0.5\parindent{$^{#1}$\hfil}\let\INS=Y\ignorespaces
\else\par\hangindent0.5\parindent\hangafter=1
\noindent\hbox to0.5\parindent{$^{#1}$\hfil}\ignorespaces\fi\fi}%
\def\mehrsterne{\advance\sterne by1\global\sterne=\sterne}%
\def\FOOTNOTE#1{\mehrsterne\ifcase\sterne
\or\bfootax \ignorespaces #1\efootax
\or\bfootay \ignorespaces #1\efootay
\or\bfootaz \ignorespaces #1\efootaz\else\fi}%
\def\PRESADD#1{\mehrsterne\ifcase\sterne
\or\bfootax Present address: #1\efootax
\or\bfootay Present address: #1\efootay
\or\bfootaz Present address: #1\efootaz\else\fi}%
\def\maketitle{\paglay%
\def\missing{ ????? }%
\setbox0=\vbox{\parskip=0pt\hsize=\fullhsize\null\vskip2truecm
\let\kka = \tamt
\edef\test{\the\MAINTITLE}%
\ifx\test\missing\MAINTITLE={MAINTITLE should be given}\fi
\aTa\ignorespaces\the\MAINTITLE\eTa
\let\kka = \tbmt
\edef\test{\the\SUBTITLE}%
\ifx\test\missing\else\aTb\ignorespaces\the\SUBTITLE\eTb\fi
\let\kka = \tams
\edef\test{\the\AUTHOR}%
\ifx\test\missing
\AUTHOR={Name(s) and initial(s) of author(s) should be given}\fi
\noindent{\bf\ignorespaces\the\AUTHOR\vskip4pt}
\let\INS=E%
\edef\test{\the\INSTITUTE}%
\ifx\test\missing
\INSTITUTE={Address(es) of author(s) should be given.}\fi
{\noindent\ignorespaces\the\INSTITUTE\vskip10pt}%
\edef\test{\the\RECDATE}%
\ifx\test\missing
\RECDATE={{\petit $[$the date should be inserted later$]$}}\fi
\edef\test{\the\ACCDATE}%
\ifx\test\missing
\ACCDATE={{\petit $[$the date should be inserted later$]$}}\fi
{\noindent Received \ignorespaces\the\RECDATE\unskip; accepted \ignorespaces
\the\ACCDATE\vskip21pt\bf S}}%
\global\fullhead=\ht0\global\advance\fullhead by\dp0
\global\advance\fullhead by10pt\global\sterne=0
{\parskip=0pt\hsize=19.5cc\null\vskip2truecm
\edef\test{\the\SENDOFF}%
\ifx\test\missing\else\insert\footins{\smallskip\noindent
{\it Send offprint requests to\/}: \ignorespaces\the\SENDOFF}\fi
\hsize=\fullhsize
\let\kka = \tamt
\edef\test{\the\MAINTITLE}%
\ifx\test\missing\message{|Your MAINTITLE is missing.}%
\MAINTITLE={MAINTITLE should be given}\fi
\aTa\ignorespaces\the\MAINTITLE\eTa
\let\kka = \tbmt
\edef\test{\the\SUBTITLE}%
\ifx\test\missing\message{|The SUBTITLE is optional.}%
\else\aTb\ignorespaces\the\SUBTITLE\eTb\fi
\let\kka = \tams
\edef\test{\the\AUTHOR}%
\ifx\test\missing\message{|Name(s) and initial(s) of author(s) missing.}%
\AUTHOR={Name(s) and initial(s) of author(s) should be given}\fi
\noindent{\bf\ignorespaces\the\AUTHOR\vskip4pt}
\let\INS=E%
\edef\test{\the\INSTITUTE}%
\ifx\test\missing\message{|Address(es) of author(s) missing.}%
\INSTITUTE={Address(es) of author(s) should be given.}\fi
{\noindent\ignorespaces\the\INSTITUTE\vskip10pt}%
\edef\test{\the\RECDATE}%
\ifx\test\missing\message{|The date of receipt should be inserted
later.}%
\RECDATE={{\petit $[$the date should be inserted later$]$}}\fi
\edef\test{\the\ACCDATE}%
\ifx\test\missing\message{|The date of acceptance should be inserted
later.}%
\ACCDATE={{\petit $[$the date should be inserted later$]$}}\fi
{\noindent Received \ignorespaces\the\RECDATE\unskip; accepted \ignorespaces
\the\ACCDATE\vskip21pt}}%
\edef\test{\the\THESAURUS}%
\ifx\test\missing\THESAURUS={missing; you have not inserted them}%
\message{|Thesaurus codes are not given.}\fi
\if M\REFEREE\let\REFEREE=Y
\normalbaselineskip=2\normalbaselineskip
\normallineskip=2\normallineskip\normalbaselines\fi
\edef\test{\the\SUMMARY}%
\ifx\test\missing\message{|Summary is missing.}%
\SUMMARY={Not yet given.}\fi
\noindent{\bf Summary. }\ignorespaces
\the\SUMMARY\vskip0.5true cm
\edef\test{\the\KEYWORDS}%
\ifx\test\missing\message{|Missing keywords.}%
\KEYWORDS={Not yet given.}\fi
\noindent{\bf Key words: }\the\KEYWORDS
\vskip3pt\line{\hrulefill}\vfill
\global\sterne=0
\catcode`\@=12}%

\MAINTITLE {Cosmic Rays}
\SUBTITLE {IV. The spectrum and chemical composition above $10^4$ GeV}
\AUTHOR {\ts Todor Stanev@1, \ts Peter L. Biermann@2, \ts Thomas K. Gaisser@1}
\INSTITUTE {@1 Bartol Research Institute, University of Delaware, Newark,
Delaware 19716, USA,
@2 Max Planck Institut f\"ur Radioastronomie, D-5300 Bonn 1, Germany}
\SENDOFF={\ts Peter L. Biermann}
\RECDATE={December 10, 1992}
\ACCDATE={March 1993}
\SUMMARY={Using the concept developed in earlier papers, that the
cosmic rays originate in three different main sites, a) the normal supernova
explosions into the interstellar medium, b) the supernova explosions into
a stellar wind, and c) powerful radio galaxies, we demonstrate in this paper
that the spectrum and chemical abundances above $10^4$ GeV can be well
understood. Using existing data on the chemical composition of cosmic rays near
TeV energies as a constraint, we adjust the parameters of the model to fit the
shower size data from the Akeno experiment; this is necessary since the
derivation  of an all particle spectrum involves an assumption about the
chemical  composition of the cosmic rays and so we have to fit the shower size
data first and then derive the all particle spectrum.  We present a successful
fit to the shower size data which allows us to draw three main conclusions: a)
For most of the energy range above $10^5$ GeV the wind explosions can account
for both chemical composition and spectrum including the knee feature, b) the
highest particle energies required from the stellar wind explosions imply a
magnetic field in  the preexisting stellar wind of at least $3$ Gauss at a
fiducial distance of $10^{14}$ cm, c) the chemical abundances above $10^5$ GeV
are dominated by  heavy nuclei such as Neon and higher.  The bump observed in
the all particle spectrum below the knee gets weakened with the proper
treatment
of the energy dependent chemical composition. At the high energy end we obtain
an  estimate of the extragalactic flux of protons.}

\KEYWORDS={Cosmic Rays -- Plasma Physics -- Supernovae -- Shockwaves}
\THESAURUS={03.10.1, 16.11.1, 19.92.1, 19.04.1}

\maketitle
\titlea {Introduction}

With the introduction of a new concept to treat particle acceleration
in shocks where the shock normal is perpendicular to the prevailing magnetic
field, Biermann (1993, paper CR I) has been able to interpret the cosmic ray
spectrum from GeV to EeV energies, including the exact spectrum and the feature
of the knee. The cosmic ray spectrum is composed of three components:
\par
1)  The explosions of normal supernovae into an approximately  homogeneous
interstellar medium drive blast waves which can accelerate to about $10^5$ GeV
for Hydrogen (Lagage and Cesarsky 1983).  For these particles the
spectrum is near $-2.75$ after taking leakage from the Galaxy into account.
Particles get accelerated continuously during the expansion of the spherical
shockwave, with the highest particle energy reached at the beginning of the
adiabatic expansion, the Sedov phase.  However, after acceleration particles
lose energy  due to adiabatic expansion and hence, the maximum particle energy
which is relevant, is the maximum particle energy achieved at the end of the
Sedov phase, when cooling sets in and the shell can break up.  At that point
the
accelerated particle population gets mixed with the interstellar medium.  We
will refer to this phase and the associated particle population as Sedov phase
explosion and Sedov phase acceleration subsequently.  We note that the
strongest
dependence of this maximum energy is on the density of the interstellar medium
with the highest particle energy being achieved in the most tenuous medium.
Protons can get accelerated up to about $10^5$ GeV in the tenuous part of the
interstellar medium.  A detailed discussion of the acceleration process and the
differences between protons, heavier nuclei and electrons as well as the
observed radio emission from normal supernova remnants is made in Biermann and
Strom (1993, paper CR III).
\par
2)  Explosions of stars into their former
stellar winds (like Wolf Rayet stars) produce particles up to energies of about
$3 \, 10^9$ GeV (for iron, and rigidity dependent); these particles have a
slightly flatter spectrum of near $-2.67$ (this is $0.08$ flatter than the
Sedov
phase explosion spectrum) up  to a rigidity dependent bend in the spectrum and
then beyond that a steeper spectrum of near $-2.97$ up to the rigidity
dependent
cutoff.  This spectral index difference is due to a difference in the detailed
acceleration efficiency, which in turn depends on particle drift energy gains.
Below the knee energy the drifts are stronger because the particles experience
a
stronger curvature drift due to turbulence, and beyond the knee energy the
particles just experience the basic curvature and gradient of the stellar wind,
since their Larmor radius gets large (the details are explained in paper CR
I).   The discussion (Biermann and Cassinelli 1993, paper CR II) of the
acceleration of particles in the strong shocks in radiosupernovae on the one
hand, and the weaker shocks in the winds of OB and WR stars on the other hand
yields the results: a) the magnetic field in the winds of WR stars is of order
$3$ Gauss at a fiducial distance of $10^{14}$ cm from the star, b) that thus
particle energies up to about $3 \,10^9$ GeV in iron nuclei are possible from
acceleration in strong supernova shocks in such winds, and c) that there is a
critical velocity of order $10^9$ cm/sec below which electron injection is
strongly inhibited.
\par
3) The hot spots of Fanaroff Riley class II radio galaxies produce particles
with
even higher energies, up to near $10^{11}$ GeV; their spectrum is approximately
$E^{-2}$ up to the pileup just below the cutoff due to the interaction with the
cosmological microwave background. This extragalactic component has been
modelled by Rachen and Biermann (1992, 1993: paper UHE CR I): this latter model
has been shown to be consistent with Fly's Eye data on the proton
contribution below EeV energies (Gaisser et al. 1993, Rachen, Stanev, Biermann
1993, paper UHE CR II).
\par
The main fraction of the cosmic rays above 10 TeV is accelerated in shocks that
traverse stellar winds. Stellar wind magnetic fields have asymptotically a
Parker type topology, with $B_\phi \sim 1/r$ dominating over most of $4
\pi$
and
a polar cap where the radial field $B_r \sim 1/r^2$ dominates. In CR I Biermann
developed a theory to treat perpendicular shocks in a spherical geometry for
particle acceleration. This process produces $E^{-7/3}$ particle spectra in the
limiting case of large shock velocities. In the geometrically small polar cap
region, however, the geometry is that of a parallel shock, which in the limit
of
large shock velocities produces $E^{-2}$ particle spectra. Although the total
amount of energy carried by the polar cap component is small, we shall show
that
it has important contribution in the region of the knee.
\par
The Parker spiral magnetic field topology requires a region near the pole where
the magnetic field is almost exactly radial; further out from the pole the
magnetic field  becomes helical to approach  asymptotically the tight
Archimedian spiral at the equator.  There is a critical angle not far from the
pole where the Larmor radii of the particles accelerated in the pole region
overlap with the Larmor radii of the particles accelerated in the Archimedian
spiral region; this critical angle separates the two regions where the
acceleration is well approximated by a shock normal parallel to the magnetic
field yielding an $E^{-2}$ spectrum for strong shocks, and the region where the
acceleration is best described by shock normal perpendicular to the magnetic
field giving an $E^{-7/3}$ spectrum.  Since the pole region where the parallel
acceleration dominates is only about $0.01$ of $4 \pi$ the contribution of this
region to the overall spectrum is very small except for the particle energies
near $10^6$ GeV, i.e. near the bend in the overall spectrum and the identical
cutoff for the polar cap particles; the polar cap particles have a steep cutoff
at the bend energy because at that energy the acceleration is limited by lack
of
space as discussed above.  Fig. 1 shows a a schematic representation of the
composite model we are discussing, broken down into its four components.
\begfig 8.8 cm
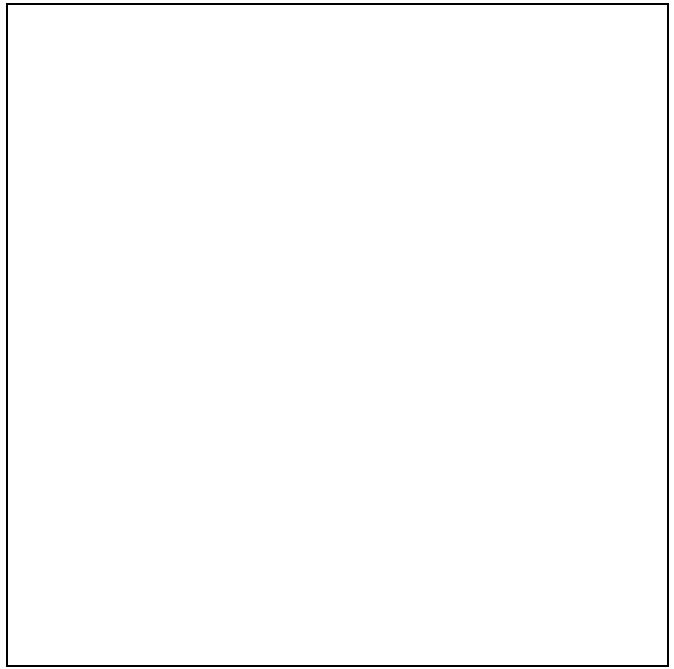
\figure{1}{The generic spectrum of one nuclear species with all
four cosmic ray components, here for Hydrogen:  Component $1$ is due to the
Sedov phase explosions in the interstellar medium, component $2$ is due to
stellar winds, with $3$ the polar cap component, and component $4$ is
extragalactic.  $E_1$ is the cutoff energy for component $1$, and $E_2$ is the
bend energy for the wind component and also the cutoff energy for the polar cap
component.}
\endfig
\par
Here we wish to test the overall model proposed by asking whether it can
successfully account for the spectrum and chemical composition at particle
energies beyond $10^4$ GeV: the special difficulty in this endeavour is the
fact
that we know the chemical abundances near TeV particle energies already, and
extrapolating these spectra does not give the bump and knee of the well
established overall particle spectrum.  On the other hand, the particle
energies
themselves and the known chemical abundances can indeed be understood with the
concept that explosions into a wind like that of a WR star are their origin
(V\"olk and Biermann 1988, Silberberg et al. 1990).
\par
The difficulty in all such attempts originates in the relation between
energy estimation with air showers and the mass of the primary nucleus.
Generally showers initiated by heavy nuclei develop and are
absorbed faster in the atmosphere. Heavy nuclei thus produce showers of
smaller size at the observation levels of all existing experiments.
The effect is stronger for inclined showers, which have to penetrate
through larger atmospheric thickness. In this paper we model the shower
size that the primary cosmic ray flux generates in the Akeno detector
(Nagano et al. 1984), using both vertical showers and slanted showers; having
performed such a fit, we then reconstruct the all particle spectrum.
\par
The paper is organized as follows: First we describe our input data and the
most important parameters; then we present the data we wish to fit (Akeno)
and the shower code used to calculate the observable parameter; our
successful fits are shown with a  discussion of possible errors in another
section; we conclude with a discussion of the newly derived all particle
spectrum and an outlook at possible next steps of interest.

\titlea {Input parameters}

Various cosmic ray experiments have given data about the chemical composition
near TeV energies and somewhat beyond, up to $100$ TeV, most notably from the
Chicago Group (Grunsfeld et al. 1988, M\"uller 1989, Swordy et al. 1990,
and M\"uller et al. 1991), the JACEE experiment (Parnell et al. 1989,
Burnett et al. 1990, and Asakimori et al. 1991a, b ), Simon et al. (1980)
and Engelmann et al. (1990). As an illustrative example we show in Fig. 2
the Oxygen data, where we have averaged existing data from all four
experiments when independent measurements exist in the same energy range.  The
data for the different experiments are quite consistent with each other.
Energy
ranges covered only by a single experiment are not represented.
\begfig 5.2 cm
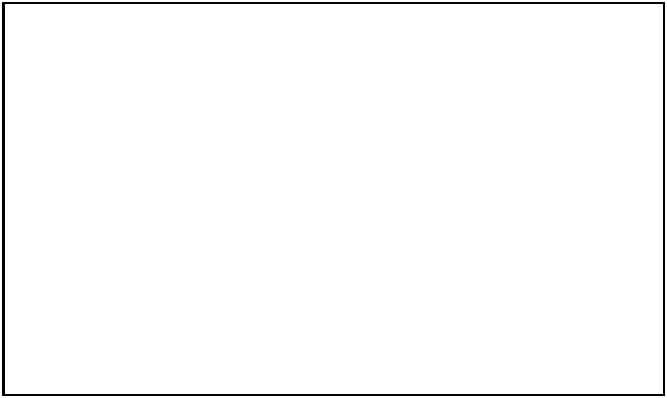
\figure{2}{The Oxygen spectrum derived from all existing data by binning
and then averaging with the weight of the error bars.  A fit is shown to the
resulting average spectrum showing a $-2.64$ spectral index.}
\endfig
Data clearly demonstrate a power law behaviour with a somewhat flatter spectrum
than the low energy index of about $2.75$. Oxygen has a spectrum of about
$2.64$,
corresponding well to the argument made in CR I, that supernova explosions into
existing stellar winds produce not only higher particle energies, but also
spectra flatter than the lower energy particles from Sedov phase explosions
into
the interstellar medium.  We use all these known chemical abundances as a given
input, combining the elements into six groups, and the sources into three
different sites as discussed above.  The six element groups which we use are:
a)
Hydrogen, b) Helium, c) Carbon, Nitrogen and Oxygen, d) Neon to Sulfur, e)
Chlorine to Manganese, and f) Iron.  We assume that the galactic cosmic rays
cut
off at $10^8 \, Z$ GeV  and that the extragalactic component is nearly all
protons.  In paper CR II (Biermann and Cassinelli 1993) we argue that this is
in fact suggested by a number of stellar observations.
\par
The main uncertain parameters are then the following:
\par
1) The spectral difference between the Sedov phase explosion particles
and the wind explosion particles.  Simple first order theory gives
$0.08$ for this parameter, the difference between $2.75$ and $2.67$, as
described above.  Arguments given in paper CR I and in Silberberg et al. (1990)
suggest that the heavier elements in the TeV range come dominantly from shock
acceleration in winds;  therefore, we ought to look for a slight difference in
spectrum between Hydrogen and heavy nuclei. For example, for Oxygen we find
$2.62 \pm 0.03$ and for Hydrogen $2.74 \pm 0.02$, which corresponds to a
spectral difference $0.12
\pm 0.04$, consistent with our theoretical
expectation
for the difference between the Sedov phase particles  and the wind shock
accelerated particles. We will therefore explore a range for the spectral index
of the wind particles below the knee close to $2.67$.
\par
2)  The spectral difference at the knee.  Again, simple theory (CR I) gives for
this number the value $0.30$, see
above.  The data on the spectrum above the knee suggest a spectral index of
$3.0
\pm 0.1$. We will try a small range for the spectral index of the particles
beyond the knee close to $3.00$.
\par
3)  The particle energy at the cutoff for the Sedov phase explosion
acceleration.  The Jacee data suggest for this number approximately $100$ TeV
for Hydrogen. Our model firmly requires that this energy scale with the charge
$Z$ of the nucleus.
\par
4)  The particle energy of the knee.  Observations suggest an energy
overall of about $5\, 10^3$ TeV.  Our theory requires that this energy also
scale with nuclear charge $Z$, and if the knee is indeed dominated by
heavy nuclei this energy for Hydrogen is likely to be above $200$ TeV.
\par
The division of the abundances between the Sedov phase accelerated particles
and the wind explosion accelerated particles can be made using existing data,
especially from JACEE.

\titlea {The data to be fitted and the shower development Monte Carlo code}

 The Akeno air shower experiment measures the size of showers arriving
at the detector at different zenith angles. The resulting shower size spectra
are converted into constant intensity curves, i.e. groups of showers
having the same experimental rate at different angles, thus initiated
by primary nuclei of the same energy.
\begfig 8.8 cm
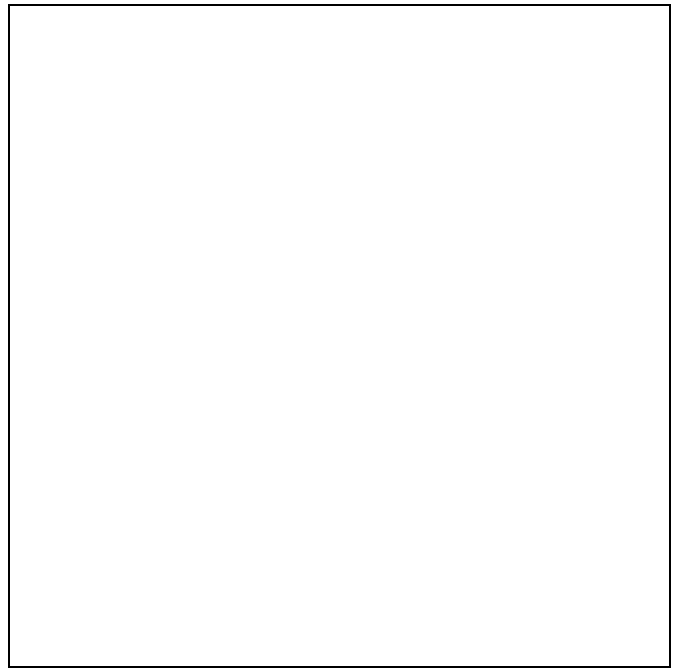
\figure{3}{Average shower profiles for a primary particle energy of $10^8$
GeV for Hydrogen, Nitrogen and Iron.  The two different slant depths which we
use are indicated.}
\endfig
The angular dependence of the constant intensity
curves is then used to extrapolate to size at shower maximum, from which
an estimate of the primary energy that corresponds to a given intensity can be
made. This fairly model independent procedure is still influenced by the
chemical
composition of the cosmic ray flux, especially if it has a strong energy
dependence.  The reason for this is illustrated in Fig. 3 which shows the
development of showers with the same primary energy but different nuclei.   We
have decided to use directly the shower size spectra to reconstruct the cosmic
ray spectrum under the assumption of different chemical compositions. We do
this
simultaneously for two zenith angles, vertical ($secan \,
\theta = 1.0$) and
for
$secan \, \theta = 1.2$, i.e. one moderate  slant angle. At these angles the
statistical errors are small. The shower size is expressed by  $N_e$ - the
total
number of charged particles at observation level. The flux of showers with
$N_e$
= $10^6$ at $secan \, \theta = 1.2$ is smaller than the vertical one by a
factor of 7. The shapes of the size spectra are also different, as seen in Fig.
4.   \par
We use a simple parametrization (Gaisser, 1979) to calculate
the size at the two slant depths (920 g/cm$^2$ for $secan \, \theta = 1.0$ and
1104 g/cm$^2$ for $secan \, \theta = 1.2$). While the original formula
is for a constant inelastic cross-section, in this estimate we use a
proton inelastic cross-section on air growing with the energy as
$(log \, E_p)^{1.8}$ (Gaisser et al., 1987). For all nuclei heavier than
Hydrogen we assume superposition, i.e. that all constituent nucleons
interact independently in the atmosphere. This assumption is known
to represent correctly the average shower size and underestimate
its fluctuations.
\par
Our technique is intermediate between numerical integration and
Monte Carlo. We step through the energy spectrum of each chemical
component in small ($10^{0.01}$) steps and calculate the shower
size generated at the two depths by a nucleus of energy E. To
estimate the fluctuations in $N_e$, which are very important in
view of the steep energy spectra, we calculate $100$ showers at each
step, sampling the interaction depth for each nucleon. The resulting
$N_e$ are binned with the appropriate weights.
\par
Although generally correct, the procedure used is not exact.  A proper
comparison
of the cosmic ray spectrum and composition model with experimental data should
account fully for the fluctuations in the shower development (i.e. use a
non-superposition nuclear fragmentation model as in Engel et al., 1992) and
those induced by the detection technique.  This is a long term project which
requires a close collaboration with the experimental groups.  The technique
used here,  which is very sensitive to the cosmic ray spectrum and composition
as we show further down, is fully sufficient as a first pass in the analysis of
air shower data in terms of rapidly changing composition.
\par
The most important point here is to remember that shower size is
reduced for particles at a given energy with higher nuclear mass: e.g. an Iron
particle and a proton of the same energy will produce very different shower
sizes, with the shower size of the Iron nucleus being much smaller (see Fig.
3).

\titlea {The model fit}

We present the fits of the shower size distributions in Fig. 4, where
the bands represent the experimental errors.
\begfig 11.3 cm
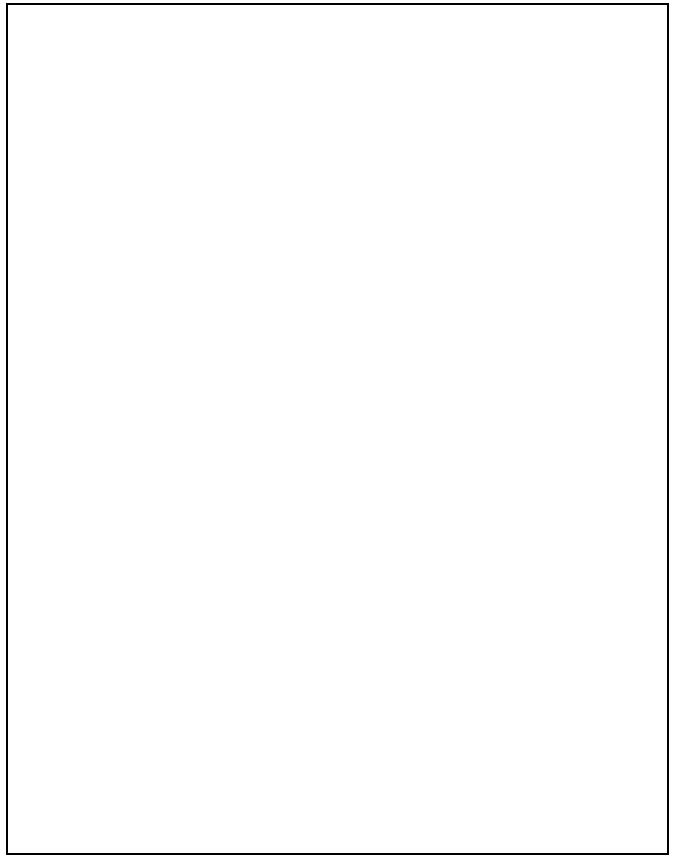
\figure{4}{The fit of our model to the Akeno shower size data, both for
vertical
showers as well as for a slanted shower.  Shown are the data as small boxes
with the shading indicating the $1$ sigma error range, as well as the model as
a bold histogram.}
\endfig
The comparison demonstrates that the fits are well within the acceptable error
ranges. The model has the following parameters:
\par
The cutoff energy (exponential cutoff) for the Sedov phase
accelerated particles is $120$ TeV for protons; the spectral index of the wind
component is $2.66$ below the knee and $3.07$ above the knee; this steepening
occurs at $700$ TeV for protons and is rigidity dependent. The ratio of the
polar cap component to the steeper wind component is $1$ for heavy nuclei at
the
bend. Although the polar cap component does not contribute to shower sizes
significantly below or above the knee, it is essential for reproducing the
sharp
break at $N_e = 10^6$.  The reproduction of this shower break, as well as the
observed continuous flattening of the spectra of all heavy nuclei requires the
introduction of a very flat ($\gamma=2$) acceleration component, i.e. the polar
cap component.  We see that all four major parameters are
close to their expected values. In Fig. 5 we show the contributions of the six
chemical composition groups which we have distinguished, H, He, CNO, Ne-S,
Cl-Mn, and Fe.   We emphasize that all three critical particle energies, the
cutoff energy for the Sedov phase accelerated particles, and both the bend
energy
and the cutoff energy for the wind accelerated particles are proportional to
the
nuclear charge Z, and are thus in our model fit $120 \,Z$ TeV, $700 \, Z$ TeV,
and $10^5 \, Z$ TeV, respectively.
\begfig 11.3 cm
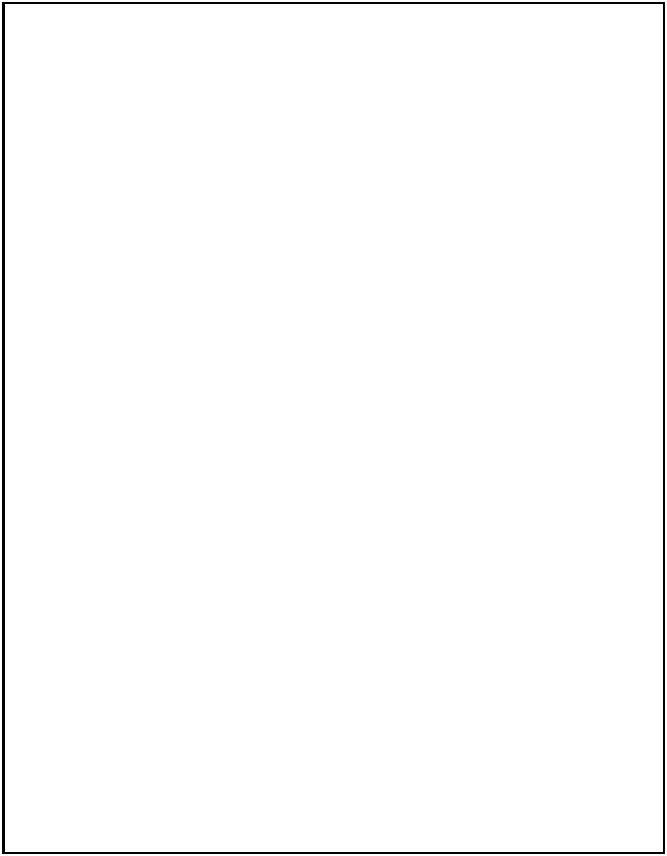
\figure{5}{The various contributions from the six element groups we
distinguish:
H (solid), He (dots), CNO (dash), Ne-S (long dash), Cl-Mn (dash dot), and Fe
(long dash dot).}
\endfig
It is clear that within the framework of our picture the knee feature in the
overall spectrum can only be due to the addition of the polar cap component.
Without it the generated size spectrum cannot have a sharp bend in any model
similar to ours, with rigidity dependent features and composed of different
nuclear components.
\par In presenting the fits we could have achieved
smoother
histograms by applying an exponential cutoff also to the polar cap component.
We
prefer to present the histograms in their present form, because a closer
inspection now allows the reader to identify the component that causes
the bend in the shower size spectra. In reality every single nuclear
component has its own rigidity dependent behaviour which of course
generates both smoother particle spectra and smoother size spectra.
\par
We note that the shower size distribution is very sensitive to changes in the
parameters. Relatively small changes (by more than $20\%$) of the cutoff
and bend energy, as well as in the composition of the wind component,
affect the calculated size spectra so strongly that they become inconsistent
with data. Together with the direct measurements in the 1 TeV region
the shower size distributions constrain the models within a small
parameter space. Fitting the shower size distribution at different zenith
angles is critical despite its fairly large error bars.
\par
We also note that the code we use for reproducing the shower size
distributions itself depends rather strongly on
model fits to accelerator data to describe high energy interactions of
nuclei; this is especially important for slanted showers where the tail of the
shower profile is measured.  Thus, the particle physics used introduces an
additional uncertainty which is difficult to assess.

\titlea {Conclusions}

In Fig. 6 we show the all particle spectrum which results from our analysis.
We compare it with the conventionally derived all particle spectrum, which does
not explicity account for changes in composition.
\begfig 8.3 cm
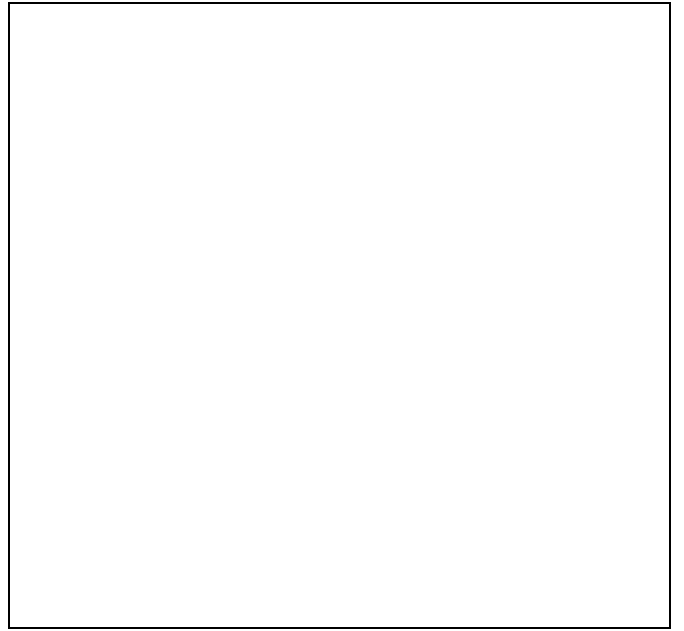
\figure{6}{Comparison of our all particle spectrum with the conventionally
derived spectrum, as well as the various contributions from the different
element groups. Here the symbols denote the following experiments:  open
circles,
Akeno; open squares, Haverah Park; open triangles, pointing down, Yakutsk;
open triangles, pointing up, Tien Shan; open hexagons, Fly's Eye; full squares,
Proton 4; full circles, Jacee.  The various references are given in detail in
Stanev (1992) and in Hillas (1984).}
\endfig
Several properties stand out:
\par
1)  The correct spectrum is lower in the knee region and beyond
than the conventionally derived spectrum; the factor is about $2$.
The knee itself is not as sharp but the change of the spectral index
takes place at approximately the same energy.
\par
2)  The composition changes rapidly in the knee region, becomes
increasingly heavier, and in our current interpretation is dominated
by the Neon group above the knee. One should note that the derivation
of the exact charge of the dominating group depends strongly on the
cascade model and that many observable parameters are not
very different for Ne- and Fe-induced showers. The data will, however, be
inconsistent with a component lighter than Ne dominating the
cosmic ray flux above 10$^7$ GeV.
\par
3)  The proton flux shown in Fig. 6 at energy above 10$^8$ Gev represents
the extragalactic component.  Because the wind component has a maximum energy
of $10^8$ GeV for protons ($2.6 \, 10^9$ GeV for Fe) the calculated shower size
spectra would become too steep for sizes above 10$^7$ without this extra
component. Although the exact shape of the extragalactic spectrum needs further
adjustments, its magnitude is limited by the comparison with the size spectra.
This implies an extragalactic flux (presumably protons) that is consistent with
the independently derived estimate by Gaisser et al. (1993) using the Fly's Eye
data; a full comparison with an extragalactic model (Rachen and Biermann, 1992,
1993:  paper UHE CR I) has been done in Rachen et al. (1993, paper UHE CR II).
\par
4)  The comparison with existing direct data on the chemical composition
is shown in Fig. 7 for the particle energy range $1$ to $100$ TeV. Hydrogen,
Helium and Iron groups data are shown.
\begfig 8.3 cm
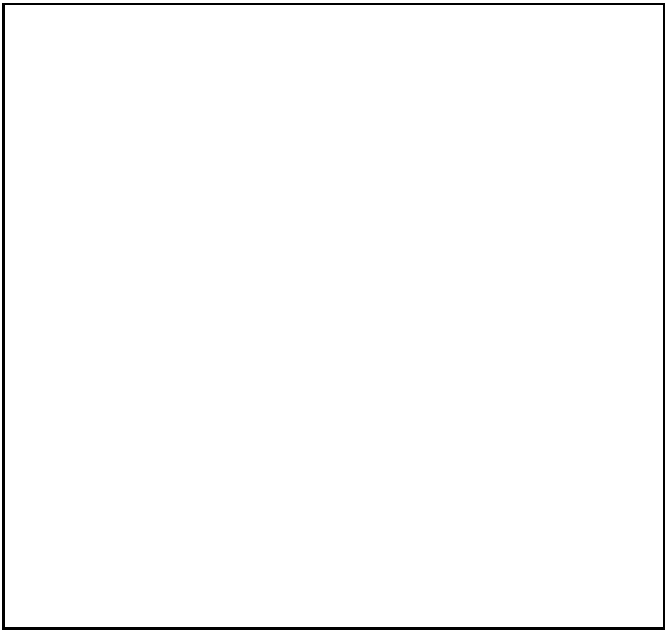
\figure{7}{A comparison of the measured element abundances for
Hydrogen, full squares, Helium, full circles, and Iron, full stars, with our
model, again together with the all particle spectrum. The sources of the data
are given in the first paragraph of section 2.}
\endfig
The comparison demonstrates that our curves fit all existing data quite well,
including the trend, observed in heavy nuclei, to exhibit a flattening of the
spectrum at the approach to the knee. The seeming inconsistency with the
highest
energy direct data for iron is possibly due to the fact the the JACEE
experiment
(where these points come from) has presented its measurements for nuclei with
$Z>17$.  It is difficult to judge what the absolute normalization of the cosmic
ray flux beyond the knee is.  The presentation of the spectrum in the form used
here ($E^{2.75} \,dN/dE$) tends to exaggerate differences in the absolute
normalization.  A relatively small error of the energy determination in air
showers (typical error of $20\%$) changes the normalization of the absolute
flux
by a large amount ($65\%$).
\par We conclude that the most stringent test yet
of
the model proposed in the earlier papers of this series (CR I, CR II, CR III,
UHE CR I, UHE CR II), which give predicted spectral shapes for
all three sites of origin for cosmic rays is successful.  Several detailed
conclusions can be made:
\par
1)  The abundances of the cosmic rays from explosions into stellar winds
do not require nor imply any admixture from the heavy elements newly
produced in the star exploding.  All the heavy elements already present
in the stellar wind prior to the explosion participate in the feeding of the
accelerated particle population, and no other additional source is
required.
\par
2)  The nuclei accelerated in the polar cap region are essential for
a successful fit of the shower size spectra. Although the total energy
carried by such nuclei is not more than $1/100$ of the total
wind component, they contribute a major fraction in the knee region.
\par
3)  There is no requirement for other cosmic ray sources, either
from spectral arguments or from abundance arguments; thus pulsars and
compact X-ray binary systems may accelerate lots of particles, but they
need not play a dominant role out in the typical interstellar medium.
\par
4)  The large fraction of heavy nuclei inferred from the Fly's Eye data
(Gaisser et al. 1993) requires the acceleration of heavy element nuclei out to
at least $3 \, 10^9$ GeV.  This implies (see CR I and II) that the winds of the
stars that explode as supernovae have magnetic fields at least as strong as $3$
Gauss at a fiducial distance from the star of $10^{14}$ cm.  Since this
limiting
particle energy is already derived by using the spatial limit given by the
condition that the Larmor radius of the particle fit into the space available,
there is no other way than indeed having these high fields in the stellar winds
of massive stars. Should a more sophisticated analysis of the
Akeno and Fly's Eye data require a galactic Cosmic Ray contribution of
iron nuclei even at $10^{10}$ GeV (see the discussion in Rachen, Stanev and
Biermann 1993, UHE CR II), then a correspondingly higher magnetic
field strength is required, with a lower limit of near $10$ Gauss at
$10^{14}$ cm from the star.
\par
The next steps possible and maybe desirable with better data or a more
complete analysis of the existing data by the experimenters are the
following:
\par
1)  One must refine the cascade code to describe more accurately
the high energy range and the highly slanted showers.  Several different valid
particle physics extrapolations to ultrahigh energy should be used for an
estimate of the errors caused by the uncertainty of the particle physics
input.  In addition, a non-superposition fragmentation model should be used
to account more correctly for the shower development fluctuations.
\par
2)  Also, one might combine
all such source based calculations with a proper propagation model in the
Galaxy;
after all, many of the heavy nuclei break up due to spallation.  The task then
would be to fit also the well known chemical abundances of the cosmic rays at
GeV energies.

\ack {The foundation for this work was laid during a five month
sabbatical in  1991 of PLB at Steward Observatory at the University of Arizona,
Tucson.  PLB wishes to thank Steward Observatory, its director, Dr. P.A.
Strittmatter, and all the local colleagues for their generous hospitality
during
this time and during many other visits.  PLB also wishes to thank Drs.
J.H. Bieging, J. Cassinelli, J.R. Jokipii, K. Mannheim, H. Meyer, R. Protheroe,
M.M. Shapiro and R.G. Strom for extensive discussions of Cosmic Ray, Supernova
and Star physics.  This work was supported by a NATO travel grant to PLB and
TS;
high energy physics with PLB is supported by the BMFT (FKZ 50 OR 9202), and DFG
Bi 191/6,7,9. Work by TKG and TS is supported by grants NSF/PHY/8915189 and
NAG5/1573.  }

\begref
\ref Asakimori et al.: 1991a in Proceedings of 22nd ICRC, (Dublin Institute for
advanced studies), Dublin, Ireland, vol. 2, p.57
\ref Asakimori et al.: 1991b in Proceedings of 22nd ICRC, (Dublin Institute for
advanced studies), Dublin, Ireland, vol. 2, p.97
\ref Biermann, P.L.: 1993, Astron.\&Astroph. (paper CR I, in press)
\ref Biermann, P.L., Cassinelli, J.P.: 1993, Astron.\&Astroph. (paper CR
     II, submitted)
\ref Biermann, P.L., Strom, R.G.: 1993, Astron.\&Astroph. (paper
     CR III, submitted)
\ref Burnett et al.: 1990 ApJ 349, L25
\ref Engel, J. et al.:  1992 Phys.Rev. D 46, 5013
\ref Engelmann, J.J., Ferrando, P., Soutoul, A., Goret, P., Juliusson, E.,
     Koch-Miramond, L., Lund, N., Masse, P., Peters, B., Petrou, N., Rasmussen,
     I.L.:  1990 Astron.\& Astroph. 233, 96
\ref Gaisser, T.K.:  1979 in Proc. Cosmic Ray Workshop, Salt Lake City, Utah
\ref Gaisser, T.K., Sukhatme, U.P., Yodh, G.B.:  1987 Phys.Rev. D 36, 1350
\ref Gaisser, T.K., Stanev, T. et al.: 1993 Phys.Rev D 47 (in press)
\ref Grunsfeld et al.: 1988 ApJ 327, L31
\ref Hillas, A.M.:  1984 Ann. Rev. Astron.\&Astroph. 22, 425
\ref Lagage, P.O., Cesarsky, C.J.:  1983 Astron.\&Astroph. 118, 223
\ref M\"uller, D.: 1989 Adv.Space Res. 9, (12)31
\ref M\"uller, D. et al.: 1991 ApJ 374, 356
\ref Nagano, M. et al.:  1984  Journal of Physics G 10, 1295
\ref Parnell, T. et al.: 1989 Adv.Space Res. 9, (12)45
\ref Rachen, J., Biermann, P.L.:  1992a in Proc. "Particle acceleration in
     cosmic plasmas", Eds. G.P. Zank, T.K. Gaisser, AIP conf.Proc. No. 264, 393

 \ref Rachen, J.P., Biermann, P.L.: 1993, Astron.\&Astroph. (paper UHE CR I,
					in press)
\ref Rachen, J.P., Stanev, T., Biermann, P.L.: 1993 Astron. \&Astroph.
     (paper UHE CR II, in press)
\ref Silberberg, R., Tsao, C.H., Shapiro, M.M., Biermann, P.L.:
     1990, ApJ 363, 265
\ref Simon et al.: 1980 ApJ 239, 712
\ref Stanev, T.:  1992 in Proc. "Particle acceleration in
     cosmic plasmas", Eds. G.P. Zank, T.K. Gaisser, AIP conf.Proc. No. 264, 379
\ref Swordy, S.P., M\"uller, D., Meyer, P., L'Heureux, J., Grunsfeld, J.M.:
     1990 ApJ 349, 625
\ref V\"olk, H.J., Biermann, P.L.:  1988 Ap.J. 333, L65
\endref
\end